\documentstyle[12pt,fleqn]{article}
\begin{document}
\baselineskip= 22 truept
\def\be{\begin{equation}}
\def\ee{\end{equation}}
\def\bea{\begin{eqnarray}}
\def\eea{\end{eqnarray}}
\begin{titlepage}
\vspace{1cm}
\begin{center}
{\large \bf Entropy of Extremal Black Holes in Two Dimensions }\\
\vspace{1cm}
{\bf Alok Kumar and Koushik Ray }\\
Institute of Physics, \\Bhubaneswar 751 005, INDIA \\
email: kumar, koushik@iopb.ernet.in \\
\today
\end{center}
\thispagestyle{empty}
\vskip 4cm
\begin{abstract}
Entropy for two dimensional extremal
black holes is computed explicitly
in a finite-space formulation of the black hole thermodynamics
and is shown to be zero {\it locally}. Our results are in
conformity with the recent one by
Hawking et al in four  dimensions.
\end{abstract}
\vfil
\end{titlepage}
\eject

It is known that black holes have
temperature and entropy and are therefore amenable to a
thermodynamic description.
For an asymptotic observer, the temperature of the black hole
is given by the surface gravity on the horizon
and the entropy is one quarter
the horizon area. It was shown by
Gibbons and Hawking\cite{GibHaw} that the entropy
of the black hole coincides with the value of the
Euclideanized action at infinity.
The laws of {\em black hole
thermodynamics} were also formulated \cite{waldb,lousto}.
It was also thought that for charged black hole,
e.g. Reissner--Nordstr{\"o}m solution, one had a
non-vanishing entropy even in the extremal limit (viz. charge
$\rightarrow$ mass) although the temperature tends to zero. This
is essentially due to the fact that extreme black holes have
non-zero area. As a result, although the extremal limit was known to be
unattainable due to the cosmic censorship\cite{nf}, there was no
purely thermodynamic way to establish this. In thermodynamics,
unattainability
of absolute zero by adiabatic processes follows from
Nernst's postulate of the vanishing of entropy in this
limit\cite{callen}.
It has been realized only recently\cite{HHR,tei} that extreme black
holes in four space-time dimensions {\em do}
have zero entropy as measured at infinity.

Recently we have studied the thermodynamics of several types of
black hole solutions in two dimensions in a finite-space
formulation\cite{kr} devised by Gibbons and Perry\cite{GP}.
It was also commented that our results applied only to
non-extremal black holes. The extremal black holes needed
separate treatment.

In this paper we apply the method of
\cite{kr} to the extremal black holes in 2D and show that the
entropy in these cases is also zero which corroborates the
results of \cite{HHR}. However in the present formulation
the local entropy itself turns out to be zero.
In this regard the extreme black holes are
thermodynamically similar to the linear dilaton vacuum
solutions. We also compute
the energy of the extreme black holes from thermodynamic
considerations and reproduce the ADM mass.
We deal with two extremal black hole solutions.
One of these is an asymptotically
flat solution for the heterotic string\cite{nappi}.
The other black hole considered
is an asymptotically anti-de Sitter one ensuing from a
charged version of the Jackiw-Teitelboim theory discussed
by Lowe and Strominger \cite{LS}.

Finite-space formulation of thermodynamics is
discussed for the non-extreme cases in \cite{GP} and \cite{kr}.
It was found that
the entropy turns out to be
non-zero on setting charge equal to mass in the
entropy for the non-extremal ones.
However, note that, in this scheme
the metric is written in a particular gauge,
viz.
\be\label{gauge}
 ds^2 = g_{tt} dt^2 + d\rho^2.
\ee
This metric is singular in the limit
charge $\rightarrow$ mass. Actually
the coordinate transformations
from Schwarzschild to these
coordinates itself is invalid in this limit.
Therefore to discuss the thermodynamics of extremal black holes
we shall change the coordinate after setting
charge equal to mass.
In the transformed coordinates, (t, $\rho$), for the extreme
cases the
horizon is at $ \rho = -\infty$ and the surface gravity at the
horizon turns out to be zero. We shall treat the temperature,
i.e. the inverse periodicity of the Euclidean time
at the horizon, as
arbitrary till the end. Thermodynamic calculations are then the
same as in \cite{GP,kr}.

Let us
now review the thermodynamic definitions in the present scheme.
We begin with the definition of free energy:
\be \label{fen}
\cal F = \frac{\cal I}{\beta},
\ee
where $\cal I $ is the Euclideanized
action evaluated for the metric of the
space-time under consideration and $\beta $ is the inverse
temperature. Therefore we will first have to evaluate
the actions for the cases at hand. Since we are
concerned with a finite-space formulation of thermodynamics,
we have to evaluate the action with the boundary
contribution  on a spacelike slice in the above mentioned coordinates.
This is feasible since the metric admits a Killing
vector $ k^a = \left( \frac{\partial}{\partial t} \right)^a $.
The inverse proper periodicity of the
Euclideanized time at a fixed value of the spatial coordinate
is interpreted as the local temperature $ T_{w} =
\beta^{-1} = T_c/\sqrt{g_{tt}}$, where $T_c$ is the inverse
periodicity at the horizon.
Note that since $T_c =0 $ in the cases at hand, the
thermodynamic quantities are to be interpreted as a limiting
value. For example,
\be
{\cal F} = \lim_{T_c \rightarrow 0} \frac{\cal I}{\beta}.
\ee
We will, for convenience of writing, not mention the limits
explicitly in the following.
The electric charge, $Q$, and the dilaton charge
\be \label{dildef}
{\cal D} = \int_{\Sigma} j
\ee
define the other thermodynamic potentials. Here
$ j_a \equiv \epsilon_{a}^{b} \nabla_b e^{-2\phi} $ is a
conserved current and
$\Sigma $ is a spacelike hypersuface.
We will therefore treat $\cal
F$ as a function of the
thermodynamic quantities $ T_{w}, Q$ and $\cal D $. Then
by the first law of thermodynamics,
one can define the entropy $ S $ and the dilaton potential
$\psi $ as:
\be \label{endef}
S  =  - \left[\frac{\partial {\cal F}}{\partial T_{w}} \right]_{\cal
D} , \,\,\,
\,\,\,\,\,\,\,\,\,\,\,\,\,\,\,\,\,\,
\psi  =  - \left[ \frac{\partial {\cal F}}{\partial {\cal D}}
\right]_{T_{w}} .
\ee
The  quantity $ {\cal E} = {\cal F} + ST_{w} $,
corresponds to the mass of the black hole,
provided it exsists, as the limiting value of the
difference between the energies for the black hole and
its asymptotic background solution
at spatial infinity.  An alternative thermodynamic scheme is
developed in \cite{pasq} by subtracting infinities in the
action. The present formulation sweeps these infinities {\em
under the carpet} as a divergent contribution to the chemical
potential $\psi$.

We now start with the discussions of the charged
2D black hole in the heterotic string theory \cite{nappi}.
These emerge as the solutions to the action:
\be \label{jt}
{\cal I} = -\int_{\cal M} \sqrt{g}
e^{-2 \phi}  \left[ R +
4(\nabla \phi )^2 + \lambda^2 - \frac{1}{2}F^2 \right] -
2\int_{\partial {\cal M}} e^{-2\phi} K,
\ee
where $K$ is the trace of the second fundamental form and $\partial
{\cal M} $ is the boundary of $\cal M$.
A black hole solution of this theory can be written as:
\bea \label{exsol1}
ds^2 & = & -{\cal G}(r) dt^2 + \frac{1}{{\cal G}(r)} dr^2
\\ {\cal G}(r) & = &1 -2me^{-\lambda r} + q^2 e^{-2\lambda r} \\
e^{-2\phi} & = & e^{-2\phi_0} e^{\lambda r} \\  \label{exsol2}
A_0 & = &  \sqrt{2} q e^{-\lambda r}.
\eea
Thermodynamics of this black hole has been discussed in
{\cite{GP}} for the non-extreme case. The
solution can be written in the form (\ref{gauge}) as:
\bea\label{metric}
ds^2 &=& - \frac{(1 - \frac{q^2}{m^2})\sinh^2 \lambda\rho}
{(1 +\sqrt{1 - \frac{q^2}{m^2}}\cosh\lambda\rho)^2
 }dt^2 + d\rho^2, \\
\phi & = & \phi_0 -\frac{1}{2}\log \left[ m + \sqrt{m^2 - q^2}
\cosh \lambda\rho \right], \\
A_0 & = & \frac{\sqrt{2} q}{ m + \sqrt{m^2 - q^2} \cosh
\lambda \rho},
\eea
and the asymptotic entropy for the black hole is
\be
S = \frac{\pi m}{\lambda} \left[ 1 + (1 - \frac{q^2}{m^2})^{1/2}
\right],
\ee
which is $\frac{\pi m}{\lambda}$  in the extremal
limit $ \frac{q^2}{m^2} \rightarrow 1$.
Another prescription  to get non-zero
asymptotic entropy in the extremal limit
is also suggested in \cite{pasq}.
However, note that, the transformation between the two
coordinates $r$ and $\rho$, viz.
\be \label{wrong}
r = \frac{1}{\lambda} \log\left(m + \sqrt{m^2 - q^2}\cosh
\lambda\rho \right)
\ee
is valid only for $ q < m $ and
the metric (\ref{metric}) is not defined in the extremal limit. For a
consistent
description, therefore, we have to start from (\ref{exsol1})
with $ q = m$ and then transform the
coordinates:
\be \label{correct}
r = \frac{1}{\lambda} \log \left( m + e^{\lambda( \rho -
\rho_0)}\right),
\ee
where $\rho_0$ is a choice of integration constant. Starting
anew with the coordinate transformation is in conformity with
\cite{HHR} where it is maintained that the extremal and non-extremal
black holes are different objects.
Note that, the extremal black hole has a different topology than the
non-extreme one. The horizon is now situated at $\rho =
-\infty$ unlike the non-extreme case, where horizon was at $\rho
= 0$.  Now both the black hole and the
asymptotic space are described using the same chart. This can be
seen from (\ref{correct}). By setting $\rho_0 = 0$, $\rho$ and $r$
in fact coincide for $m = 0$. This
is of import in the calculation of the mass of the black hole.
The transformed extremal solution takes the form,
\bea
ds^2 & = & - {\cal G(\rho)}dt^2 + d\rho^2, \\
{\cal G(\rho)} & = & \frac{e^{2 \lambda (\rho - \rho_0)}}{(m +
e^{\lambda (\rho - \rho_0)})^2} \\
\phi & = & \phi_0 -\frac{1}{2} \log [(m + e^{\lambda (\rho -
\rho_0)})] \\
A_0 & = & \frac{\sqrt{2}m}{m + e^{\lambda (\rho - \rho_0)}}.
\eea
We have kept the constant $\rho_0$ in the above expressions. We
will however see later on that thermodynamics is independent of
$\rho_0$.
The extreme black hole also asymptotes to the flat space
with linear dilaton
\bea
ds^2 & = &- dt^2 + d\rho^2, \\
\phi & = & \phi_0 -\frac{1}{2} \lambda (\rho - \rho_0).
\eea
We now evaluate the Euclidean action for the black
hole (\ref{exsol1})--(\ref{exsol2}). It takes the form
\be \label{act}
{\cal I} = -\int_{\partial{\cal M}} \sqrt{\frac{1}{g_{11}}}
e^{-2\phi} \left(
\frac{1}{2} \frac{\partial_1
g_{00}}{g_{00}} - 2 \partial_1 \phi\right)
\ee
The dilaton charge is found to be
\be
{\cal D} =  e^{-2\phi_0} (m + e^{x}) , \,\,\,\,\,\,\,\,\,\,\,\,
\,\, {\rm where} \,\,\,\,\,\,
x = \lambda (\rho - \rho_0).
\ee
The free energy of this black hole is
then obtained by dividing the action by the local temperature
and takes the form
\be \label{fey}
{\cal F} = -2 \lambda {\cal D}.
\ee
We note that the free energy is dependent on only one
thermodynamic quantity, namely the dilaton charge.
Hence the
entropy as defined in (\ref{endef}) is identically zero.
It also implies that the thermal energy
${\cal E} = {\cal F} = -2\lambda \cal D$.
Calculating the free energy in the similar way for the
flat space linear dilaton vacuum one finds
\bea
{\cal D_{\rm fs}} & = & e^{-2\phi_0} e^{x} \\
{\cal F} & = & -2\lambda {\cal D_{\rm fs}} \\
S & = & 0 \\
{\cal E}_{\rm fs} & = & -2\lambda {\cal D_{\rm fs}}
\eea
Hence the mass of the black hole can be obtained as the
asymptotic value of the energy differene of the two space-times
as
\be
M = \lim_{x \rightarrow \infty }
({\cal E - E_{\rm  fs}}) = me^{-2\phi_0},
\ee
which is in fact the ADM mass of the black hole.
Note that unlike the non-extremal cases \cite{GP,kr}, there is
no change of variable involved in calculating $M$
due to their description by the same charts in
this case. Interestingly, we also see the similarity in the
expressions for free energies of the black hole and the linear
dilaton vacuum solutions.

Now let us turn to the study of the black hole obtained in
\cite{LS} by introducing gauge fields
through the dimensional compactification
of a three dimensional string effective action. The
thermodynamics for the non-extremal case was discussed in
\cite{kr}.
The 2D action in this case has the form,
\be \label{action32d}
{\cal I} = -\int_{\cal M_{\rm 2}} \sqrt{g} e^{-2 \Phi} [ R +
 2 \lambda^2 -\frac{1}{4}e^{-4\Phi} F^2] - 2\int_{\partial
{\cal M}} e^{-2\Phi} K,
\ee
where now $ \Phi $ is a scalar field coming from the
compactification and plays the role of  dilaton for the 2D
action. Action
(\ref{action32d}) describes the Jackiw-Teitelboim theory with a gauge
field.
This theory possess the black hole solution
\bea \label{lsbh1}
ds^2 & = & -( \lambda^2 r^2 - m + \frac{J^2}{4 r^2}) dt^2 +
( \lambda^2 r^2 - m + \frac{J^2}{4 r^2} )^{-1} dr^2, \\ \label{lsbh}
A_0 &= &- \frac{J}{2 r^2}, \\ \label{lsbh2}
e^{- 2 \Phi} &= &r.
\eea
The parameter $J$ in this solution gives charge to this
black hole.
The metric has a curvature singularity at $ r = 0 $
for nonvanishing $J$ as is seen from the Ricci scalar
\be
R = - 2 \lambda^2 - \frac{ 3 J^2 }{2 r^4}.
\ee
It goes asymptotically, $r \rightarrow \infty$, to the
anti-de Sitter Space-time.
Now we study the thermodynamics of these black
hole solutions in the extremal case.
The use of the equations of motion
implies the following value of the classical action\cite{kr}:
\be \label{lsact}
{\cal I} = -\int_{\partial{\cal M}} \left[ n^a F_{a b} A^b
e^{-6\Phi} + 2 K e^{-2 \Phi} \right].
\ee
We now rewrite the solution
(\ref{lsbh1})--(\ref{lsbh2})
in the extremal case, $ m^2 = \lambda^2 J^2 $,
in the form (\ref{gauge}):
\be \label{transf}
r^2 = \frac{m}{2\lambda^2}\left[ 1 + e^{2 \lambda\rho} \right].
\ee
We find
\be
ds^2 = -{\cal G}(\rho) dt^2 + d\rho^2, \\
\ee
where
\be
{\cal G}(\rho) = \left( \frac{m}{2} \right) \frac{e^{4
 \lambda \rho}}{1 + e^{2\lambda\rho}}.
\ee
$A_0$ and $e^{-2\Phi}$ are still given by
(\ref{lsbh})--(\ref{lsbh2}) with $r$ replaced from (\ref{transf}).
In these coordinates the horizon is again
at $\rho = -\infty$. The free energy is
obtained by the evaluation of (\ref{lsact}). We note
however that there is
an ambiguity in the evaluation of (\ref{lsact})
due to the freedom of a
constant shift in the gauge potential: $A_a\rightarrow A_a +
const.$, in the equations of motion. Constant shifts
on gauge potentials have been
applied earlier\cite{KOP,GibHaw}.
This shift was also needed in \cite{kr} for
consistency of the present method of computations
with the Noether's charge prescription \cite{wald}.

The dilaton charge $\cal D$ is now given by
\be \label{lsdil}
{\cal D} = \left[ \frac{m}{2\lambda^2} (1 + e^{2x})
\right]^{1/2}, \,\,\,\,\,\,\,  {\rm where}
\,\,\,\,\,\, x = \lambda \rho.
\ee
Evaluating (\ref{lsact}), with a shift in the gauge
potential $ A_\mu \rightarrow A_\mu(\rho) - A_\mu (\rho =
-\infty ) $ the free energy once again takes the form
\be \label{lsfe}
{\cal F} = -2 \lambda {\cal D}.
\ee
Hence the entropy is zero again.
The thermal energy of the black hole is
to be computed with reference to the AdS linear dilaton vacuum
defined by
\bea \label{ads}
ds^2 & = & - \frac{m}{2}e^{2\lambda\rho} dt^2 + d \rho^2 ,\\
\Phi & = & -\frac{1}{2}\log \sqrt{\frac{m}{2\lambda^2}}
  -\frac{1}{2}\lambda\rho.
\eea
The free energy becomes
$
{\cal F}_{AdS} = -2\lambda {\cal D}_{AdS},
$
which implies $ S_{AdS} = 0 $ and $ {\cal E}_{AdS} = -2\lambda
{\cal D} $. Then defining $ M \equiv {\cal E} - {\cal
E}_{AdS} $ we obtain
\be
M = \left(\frac{m}{2}\right)^{1/2}e^{-x}
\ee
This gives the correct ADM mass of the black hole,
$\frac{m}{2}$, by taking into account the
redshift factor.

To conclude, we have shown that the entropy for the extremal black
holes vanishes in two dimensions.
We in fact found that the entropy
vanishes locally as could be expected from the vanishing of the
local temperature. It will be intersting to understand these
results from the microscopic point of view. Also,
two dimensional heterotic string theory have another type of
charged black hole solution\cite{rab}. It will be interesting to
investigate the thermodynamics of these solutions.

\end{document}